\documentclass{acm_proc_article-sp}
\pdfoutput=1 

\usepackage[utf8]{inputenc}
\usepackage[T1]{fontenc}
\usepackage{graphicx}
\usepackage{placeins} 
\usepackage{multirow}
\usepackage{fancybox}
\usepackage{subcaption}
\usepackage{listings}

\usepackage[english]{babel}

\usepackage{array}

\newcolumntype{C}[1]{>{\centering}m{#1}}

\title{Transaction Level Analysis for a Clustered and Hardware-Enhanced Task Manager on Homogeneous Many-Core Systems}

\numberofauthors{1} 

\author{
\alignauthor Daniel Gregorek, Robert Schmidt, Alberto Garc\'{i}a-Ortiz\\
\affaddr{Institute of Electrodynamics and Microelectronics, University of Bremen, Germany}\\
\email{\{gregorek,agarcia\}@item.uni-bremen.de, r.schmidt@uni-bremen.de}
}

\toappear{
\hrule
\vspace*{12mm}
Space reserved for the copyright notice  \\ 
Submitted to HIP3ES 2015 workshop \\ 
\vspace*{12mm}
\hrule
}

\makeatletter
\def\@copyrightspace{\relax}
\makeatother
\begin{document}

\maketitle

\begin{abstract}\label{sec:abstr} 
The increasing parallelism of many-core systems demands for efficient
strategies for the run-time system management. Due to the large number
of cores the management overhead has a rising impact to the overall
system performance. This work analyzes a clustered infrastructure of
dedicated hardware nodes to manage a homogeneous many-core
system. The hardware nodes implement a message passing protocol and
perform the task mapping and synchronization at run-time. To make
meaningful mapping decisions, the global management nodes employ a
workload status communication mechanism.

This paper discusses the design-space of the dedicated infrastructure
by means of task mapping use-cases and a parallel benchmark including
application-interference. We evaluate the architecture in terms of
application speedup and analyze the mechanism for the status
communication. A comparison versus centralized and fully-distributed
configurations demonstrates the reduction of the computation and
communication management overhead for our approach.

\end{abstract}


\terms{design, architecture}
\keywords{many-core, embedded system, run-time management, message passing, task mapping, dedicated hardware}

\section{Introduction}
\label{sec:intro}

Power-efficiency and scalability has been a driver for a variety of
cluster-based many-core systems. Among them, the P2012 (a.k.a. STHORM)
many-core architecture, the MPPA manycore and the Single-Chip Cloud
Computer SCC have recently been implemented as real-world hardware
instances \cite{benini2012p2012} \cite{dinechin2013clustered}
\cite{howard201048}. Their designs address power budgets ranging from
2W to 125W and incorporate a multitude of architectural features and
programming models.

The domain of many-cores leads to the demand for a sophisticated
\mbox{(re-)design} of the run-time task management. The task
management has to bring the dynamic requirements of the user
applications into accordance with the monitored state of the
chip. Also, a task manager is responsible for allocating the resources
(1) computation, (2) communication and (3) memory to the applications.
Hardware-assistance has become a key factor to reduce the overhead
introduced by the run-time task manager \cite{nollet2010safari}.

The idea of hardware task scheduling can be tracked back to the POLYP
mainframe computer \cite{manner1984hardware}. An overview about
separate task synchronization subsystems is given by Herkersdorf
\cite{herkersdorf2011hardware}. But to our best knowledge, we are the
first to present and to analyze a \textit{full-fledged} on-chip task
management infrastructure using a dedicated infrastructure of hardware
nodes. Key objective of the dedicated infrastructure is to conceal the
resulting management overhead from the user tasks.

The remainder of this paper is organized as follows: \mbox{Section
  \ref{sec:rel}} discusses related work. In Section \ref{sec:arch} we
present our proposed architecture and Section \ref{sec:tman}
introduces the run-time task manager. Section \ref{sec:exp} shows
experimental results, and finally Section \ref{sec:sum} concludes the
paper.

\section{Related Work}
\label{sec:rel}

A hardware-assisted run-time software for embedded many-cores is
presented by HARS \cite{lhuillier2014hars}. But, while using the
hardware sema\-phores included in the STHORM many-core architecture
their evaluation is limited to intra-cluster task synchronization.

A distributed run-time is proposed for the MPPA
\cite{dinechin2013distributed}. The run-time environment exploits a
dedicated system core which acts as a resource manager inside a single
cluster. However, their approach is constrained to a compile-time
(static) mapping scheme.

The SCC comes with a default Linux configuration and the message
passing programming model. Also, basic synchronization primitives are
implemented in hardware \cite{reble2012evaluation}. The SCC consists
of small-size clusters which yet not contain a dedicated management
core.

Besides clustered solutions there exist centralized as well as
fully-distributed approaches. Nexus++ uses a single application
specific circuit resolving time-critical task dependencies at run-time
\cite{dallou2012hardware} and applies a trace-based description of a
H.264 benchmark.  A distributed and dedicated hardware approach has
been implemented by \mbox{Isonet \cite{lee2013isonet}}. Isonet applies
a fully-distributed network of dedicated management nodes for hardware
supported load balancing.

\section{System architecture}
\label{sec:arch}

\begin{figure}[h]
\centering
\includegraphics[width=.48\textwidth]{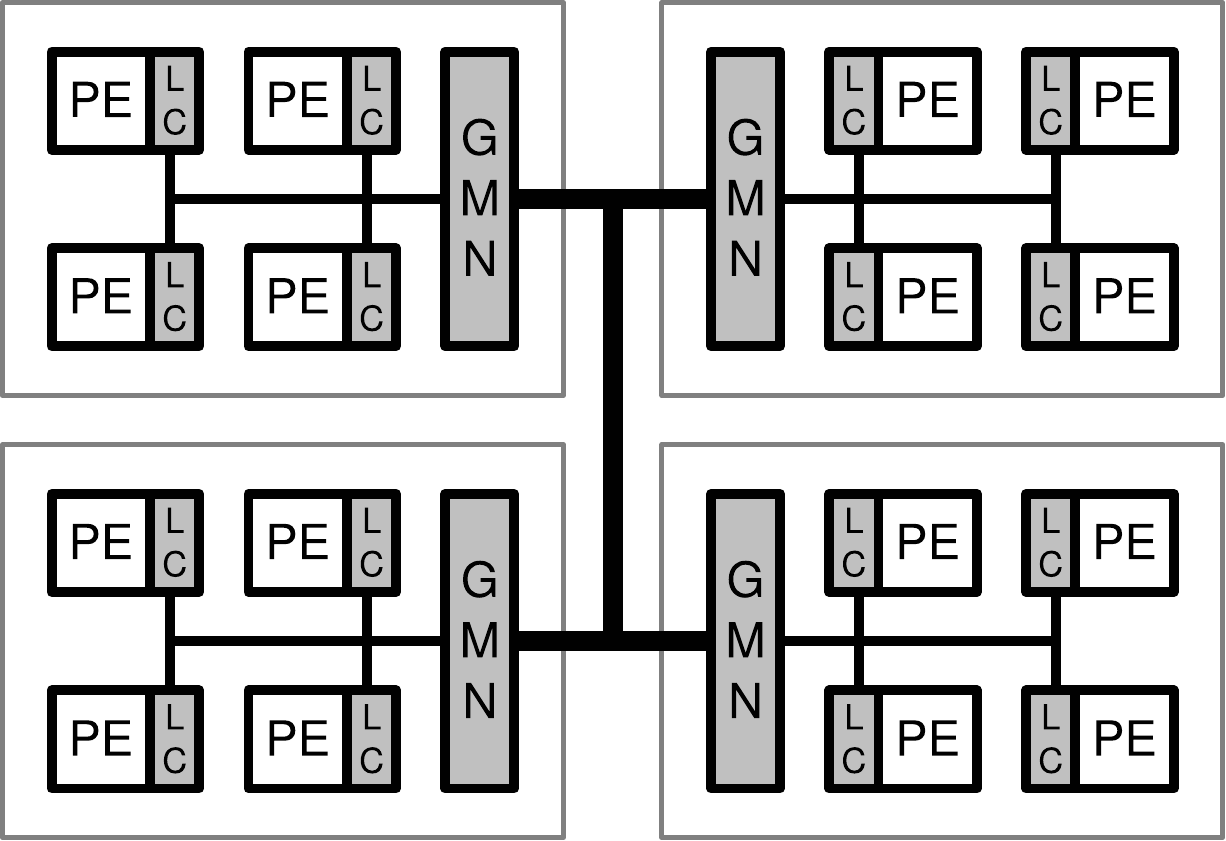}
\caption{ Outline of the system architecture for the clustered task
  management. Having $k=4$ global management nodes (GMN) and $m=16$
  local controllers (LC) coupled to the processing elements.}
\label{fig:arch}
\end{figure}

This paper is a continuation of our work presented in
\cite{gregorek2014transaction} and analyzes a clustered architecture
for the task management. Our overall system architecture is
constructed by a \textit{homogeneous} many-core system as a baseline
which is enhanced by a dedicated management infrastructure.  The
dedicated management infrastructure is implemented as a network of
global management nodes and clusters of local controllers. Each local
controller is tightly coupled to a processing element. The global
management nodes are connected to a global interconnect. A local
interconnect links one global management node with its local
controllers. \mbox{Fig. \ref{fig:arch}} gives an outline of the
proposed architecture. The interconnects are implemented as (but not
restricted to) shared buses. A common interconnect between the
processing elements is left out for better readability.  The
communication between the dedicated nodes is done by means of message
passing. Each node contains message queues for transmission and
reception.

\bigskip

\medskip

\subsection{Global Management Nodes}

Each of the global management nodes runs one instance of the run-time
task manager in software and contains dedicated hardware for message
processing. The communication between the nodes is determined by the
message protocol explained in Sec. \ref{sub:mssg}. The execution of
the system-calls from the user tasks is realized by the global
nodes. Additionally, they implement a hierarchical task mapping
algorithm and a cluster status communication mechanism at run-time
(see Sec. \ref{sec:tman}).

The global management nodes demand for programmability and for a
minimal area footprint. Messaging between the nodes requires for fast
interrupt handling.  We plan to implement the global management nodes
by means of programmable stack machines. Stack machines have very-low
hardware \mbox{complexity \cite{leong1998fpga}}, exhibit high
performance in subroutine calls (context switching), and achieve
deterministic time for interrupt handling \cite{Koopman1989}
\cite{Koopman1990}. Another advantage is the small code size of
programs written for stack \mbox{machines \cite{Bowman2010}}. Small
memory footprints allow to spend each global node its own program
memory, which diminishes communication overhead for instruction
fetching.

\subsection{Local Controller}

A local controller (LC) is tightly coupled to a processing element
(PE) for user task execution. The PE contains a functional model of a
RISC-like processor architecture and executes a trace-based
description language. The traces are used to raise the system-calls
given in Tab. \ref{tab:sysc} and determine the application behavior
(see Sec. \ref{sec:exp}).

The local controller maintains a system call dispatcher for
low-latency response and has access to the PE registers.  The
dedicated LC can be implemented with low area \mbox{overhead
  \cite{gregorek2013scalable}} and operates in parallel to the PE. Any
system-call from a user task is fetched by the LC and forwarded to its
global node by means of a dedicated message. Due to the dedicated
infrastructure for the task management the PE does only execute the
user tasks.

\subsection{Messaging Protocol}
\label{sub:mssg}

We send messages via the dedicated interconnects for communication
between the hardware nodes. The message passing combines data
transport and run-time system synchronization. Each message has a
header and one or more 32-Bit data fields. Table \ref{tab:head}
displays the structure of a message. The header contains the message
type, at least the source address, the priority and a broadcast
flag. The size of the message header depends on the actual hardware
configuration (i.e. number of nodes / address-width).

\begin{table}[h]
\caption{Message structure}
\label{tab:head}
\centering
{\ttfamily
\begin{tabular}{||p{8mm}|p{6mm}|p{6mm}|p{8mm}|p{8mm}||C{16mm}||}
\hline
type & src & dst & prio & flag & data \tabularnewline
\hline
\end{tabular}
}
\end{table}

Most of the message types directly correspond to the system calls
given in Tab. \ref{tab:sysc} and are send from a local controller to
its global node. Beyond that, a message \texttt{task-start} invokes
the start of a task. That message transports the address of the
\texttt{task-control-block} (see \cite{labrosse1998microc}) and the
\texttt{stack-pointer} as message data, and can be send from a global
node to a local controller as well as to another global node. Further,
the global nodes use the message \texttt{status-beacon} to broadcast
the current workload status (see Sec. \ref{sec:stat}) to all other
global nodes.

\section{Task Manager}
\label{sec:tman}

The task manager we use is loosely based on the Micro-C/OS-II
  \cite{labrosse1998microc} software operating system. We do not
adapt real-time capabilities but extended the task manager to have
basic multi-core functionality. The extensions to the task manager are
explained in Sec. \ref{sec:mapp} and \ref{sec:stat}. We replaced the
task scheduler to employ a simple first-come-first-serve strategy.

The system-calls, which we apply throughout this paper are explained
in \mbox{Tab. \ref{tab:sysc}}. We use a customized join/barrier
mechanism to synchronize the user tasks. To reduce the number of
system-calls, a child task is allowed to exit immediately, when
signalizing a \texttt{join-exit}.

\begin{table}[h]
\centering
\caption{ System calls }
\label{tab:sysc}
\begin{tabular}{|p{15mm}||p{9mm}|p{48mm}|}
\hline
Name               & Param.          & Description \tabularnewline
\hline\hline
\texttt{rcsv-spwn} & \texttt{imem}, \texttt{dmem}, \texttt{cnt} & Spawn new recursive task of given count (\texttt{cnt}) and instruction- (\texttt{imem}) and data (\texttt{dmem}) memory addresses\tabularnewline
\hline
\texttt{rcsv-exit} & \texttt{addr}   & Terminate task                                \tabularnewline
\hline
\texttt{join-init} & \texttt{cnt} & Initialize join barrier with given initial count and return its address to user\tabularnewline
\hline
\texttt{join-free} & \texttt{addr} & Free join barrier from memory               \tabularnewline
\hline
\texttt{join-wait} & \texttt{addr} & Let task wait until counter is zero      \tabularnewline
\hline
\texttt{join-exit} & \texttt{addr} & Decrement counter and terminate task     \tabularnewline
\hline
\end{tabular}
\end{table}

\subsection{Task Mapping}
\label{sec:mapp}

The task mapping algorithm is part of the software OS and is
implemented inside every global node. Since our targeted task
scheduling problems consist of task sets having a large number of
tasks, we use a recursive task spawning/fork strategy. Every recursive
task spawns two additional helper task and then blocks until its
child's have terminated. The recursion is executed until one of the
following stop conditions is reached:

\begin{enumerate}
\item 
The number of remaining child tasks is smaller than or equal to the
number of PEs per cluster
\item 
The number of active helper tasks is greater than or equal to the number of clusters
\end{enumerate}

The recursive start-up follows a dynamic cluster mapping procedure,
which tries to equally distribute the recursive helper tasks onto the
clusters. After the binary fork-tree has stopped to expand, the actual
child tasks of the application are spawned. This final number of
working child tasks is fixed and determined by the application
profile.

The mapping problem is therefore \textit{split} into two stages: At
the first stage, the mapping algorithm is responsible for selecting
the global nodes (clusters), where the helper tasks get mapped to. At
the second stage, the mapping algorithm selects the local processing
elements, where the actual child tasks get mapped to. Each single
mapping decision is done by means of a min-search. The mapping
algorithm chooses that node with the minimal number of mapped
tasks. To do this, every global node maintains a data structure about
the \mbox{per-PE} workload inside his private cluster and a data
structure about the summarized workload for each remote global
node. In the current implementation, we estimate the workload by
counting the total sum of locally mapped tasks.

Mapping is done only once, we do not allow a task to restart at any
different location (run-time migration), since these operations
usually come at a high performance \mbox{penalty
  \cite{bertozzi2006supporting}} and are not in the focus of our
analysis.

\subsection{Status communication}
\label{sec:stat}

Communicating the workload status is required for allowing the mapping
algorithm to make meaningful decisions.  Due to the shared nature of
the global bus interconnect we use a broadcast message to inform all
collaborating nodes about the local workload. We implemented a
threshold-based mechanisms for broadcasting the total sum of locally
mapped tasks. The mechanism triggers a broadcast, every time a certain
threshold $\Delta n_{th}$ in change of the number of mapped tasks is
reached.

\section{Evaluation}
\label{sec:exp}

We employ a simplified task-based programming model for our
analysis. Parent tasks may spawn numerous child tasks and wait until
their computation has finished. Our main criterion for evaluation is
the throughput \mbox{time $t_{r}$} (response time) of the overall
application (parent + childs). We measure the speedup as the ratio of
the sequential throughput time $t_{r,seq}$ vs. the achievable parallel
throughput time $t_{r,par}$ and show that the achievable speedup
\mbox{$S=t_{r,seq}/t_{r,par}$} is either limited by the computation or
communication management overhead.

\subsection{Analytic Model:} 

Having $n$ independent child tasks of equal length $l$, $m$
homogeneous processing elements and $k$ global management nodes the
maximal achievable speedup is limited by a temporal management
overhead $\Omega(m, n, k)$ as shown in \mbox{Eqn. (\ref{eqn:psup})}:

\begin{equation}
\label{eqn:psup}
S=\frac{t_{r,seq}}{t_{r,par}} = \frac{n\cdot l}{t_{r,par}}=\frac{n\cdot l}{ \lceil {n/m} \rceil \cdot l + \Omega(m,n,k)}
\end{equation}

\begin{figure*}
\centering
\begin{subfigure}[c]{0.49\textwidth}
\centering
\includegraphics[width=0.96\textwidth]{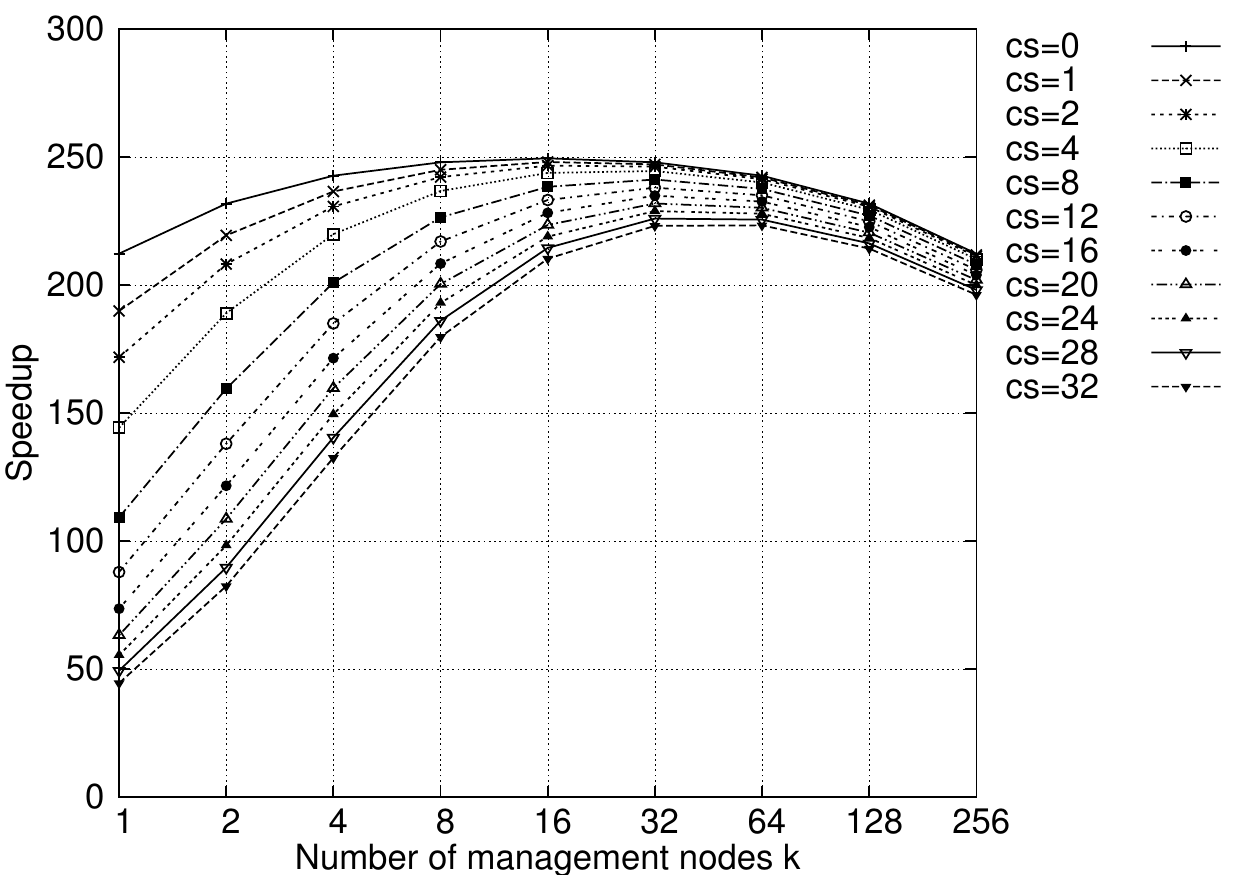}
\captionsetup{width=0.8\textwidth}
\caption{ Analytic model for the speedup using the recursive task
  startup. Having $m=256$ PEs and $n=256$ child tasks for a varying
  number of global nodes $k$ and coefficient $c_s$}
\label{fig:mod}
\end{subfigure}
\begin{subfigure}[c]{0.49\textwidth}
\centering
\includegraphics[width=0.96\textwidth]{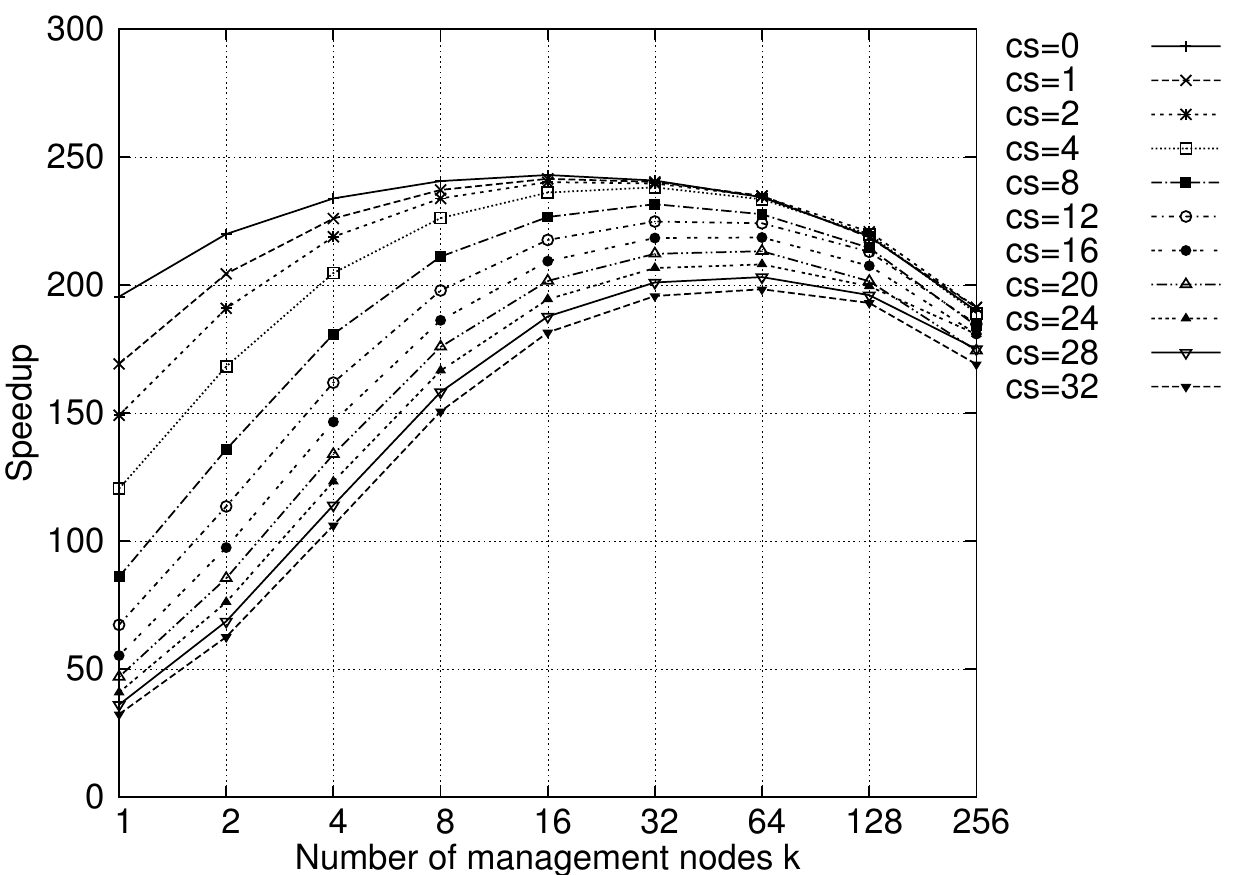}
\captionsetup{width=0.8\textwidth}
\subcaption{ Measured result for the speedup using the recursive task
  startup. Having $m=256$ PEs and $n=256$ child tasks for varying
  global nodes $k$ and the delay coefficient $c_s$}
\label{fig:meas}
\end{subfigure}
\caption{Independent tasks on 256 homogeneous processing elements}
\hrule
\end{figure*}

Due to the considered run-time computation of the mapping problem
there is a computation overhead $\Omega_{cmp}$.  Having multiple
global nodes $k$ there is an overhead $\Omega_{msg}$ in communication. We
constitute the overall management overhead $\Omega$ depending on the
number of processing elements $m$, the number of user tasks $n$ and
the number of global management nodes $k$ by equation (\ref{eqn:over}):

\begin{equation}
\label{eqn:over}
\Omega(m,n,k) =  \Omega_{cmp}(m,n,k) + \Omega_{msg}(m,n,k)
\end{equation}

Each decision of our task mapping algorithm (See Sec. \ref{sec:mapp})
infers a selection time overhead $\Omega_s$. Due to the recursive task
startup there is a logarithmic dependency ($\log n$) for the global
mapping stage. The resulting overhead $\Omega_{cmp}$ for computing
the mapping problem of $n$ user tasks is given by equation
(\ref{eqn:comp}):

\begin{equation}
\label{eqn:comp}
\Omega_{cmp}(m,n,k) =  \overbrace{\log (n) \cdot \Omega_{s}(k)}^{map~global} + \overbrace{\frac{n}{k} \cdot \Omega_{s}\left(\frac{m}{k}\right)}^{map~local}
\end{equation}

The required search function for the mapping algorithm can be
implemented having logarithmic time-complexity $\mathcal O(\log \nu)$
by e.g. Red-Black \mbox{Trees \cite{cormen2001introduction}}. The
selection time $\Omega_s$ for one decision of the mapping is modeled
as $\Omega_s = c_s \cdot \log \nu$; where $\nu$ is the number of nodes
to be searched through and $c_s$ is a timing parameter of our
framework (see Tab. \ref{tab:dflt}). Correspondingly, the
communication overhead due to intra- and inter-cluster messaging is
approximated by means of \mbox{Eqn. \ref{eqn:sync}}:

\begin{equation}
\label{eqn:sync}
\Omega_{msg}(m,n,k) = \overbrace{c_b \cdot k}^{global} + \overbrace{c_b \cdot \frac{m}{k}}^{local}
\end{equation}

Eqn. \ref{eqn:sync} introduces the timing parameter $c_{b}$ to model
the time delay inquired by communication messages. In
Fig. \ref{fig:mod} the projected speedup is plotted for the analytic
model. We set $m=256$ PEs and $n=256$ child tasks while varying the
number of global nodes and the coefficient $c_s$. As indicated, the
recursive startup and task mapping favors a number of $32 - 64$ global
management nodes.

\subsection{Experimental Setup}
We use the transaction-level simulator presented in
\cite{gregorek2014transaction} to evaluate our architecture and to
compare the analytic model against the simulation result. \mbox{Table
  \ref{tab:dflt}} gives the default parameters for our model.  Our
evaluation ignores wire capacitances, which factual privileges
fully-centralized or fully-distributed configurations with a large
number of nodes attached to the local or global interconnects. To
eliminate the effect of bottlenecks at the interconnects we previously
analyzed and set the bit-width of the buses to a convenient value of 
 \mbox{32 bit}.

\begin{table}
\centering
\caption{Default parameters for the analytic evaluation and the transaction level simulations}
\label{tab:dflt}
\begin{tabular}{|l||l|}
\hline
Name                                  & Value \\
\hline\hline
Number of processing elements         & 256 \\
\hline
Global bus width                      & 32 bit\\
\hline
Local bus width                       & 32 bit\\
\hline
Message receive delay ($c_b / 2$)     & 4 Ticks \\
\hline
Message transmit delay ($c_b / 2$)    & 4 Ticks \\
\hline
Selection delay coefficient ($c_{s}$) & 8 Ticks\\
\hline
Max. child task length                  & 16000 Ticks \\
\hline
Simulation length                     & 1e7 Ticks\\
\hline
\end{tabular}
\end{table}

\subsection{Independent Tasks}
The benchmarks are modeled by means of a trace description
language. The traces describe the computation and memory access
patterns of the tasks as well as the calls to the run-time services
(system calls).  The traces are interpreted and executed by the model
for the processing elements. In our current analysis we use a
synthetic parallel benchmark consisting of $n$ independent tasks
without any memory access. Fig. \ref{fig:meas} shows the measured
speedup fitting quite well to the analytic description due to the
regular nature of the benchmark.

\begin{figure*}
\centering
\begin{subfigure}[c]{0.49\textwidth}
\includegraphics[width=.96\textwidth]{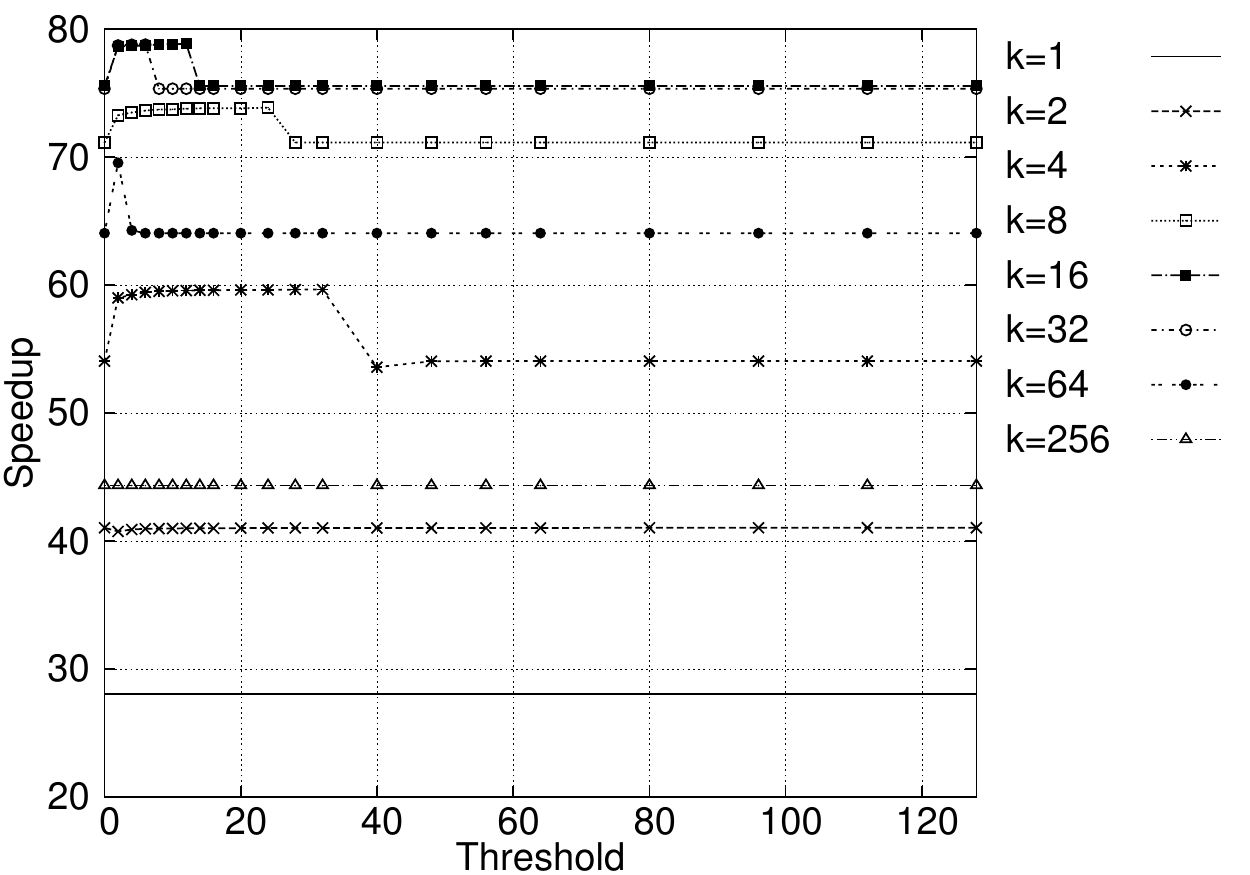}
\captionsetup{width=0.88\textwidth}
\caption{ Evaluation of the speedup versus the \mbox{threshold $\Delta
    n_{th}$} for the threshold-based workload status communication
  mechanism having different numbers of global nodes}
\label{fig:thr}
\end{subfigure}
\begin{subfigure}[c]{0.49\textwidth}
\centering
\includegraphics[width=.96\textwidth]{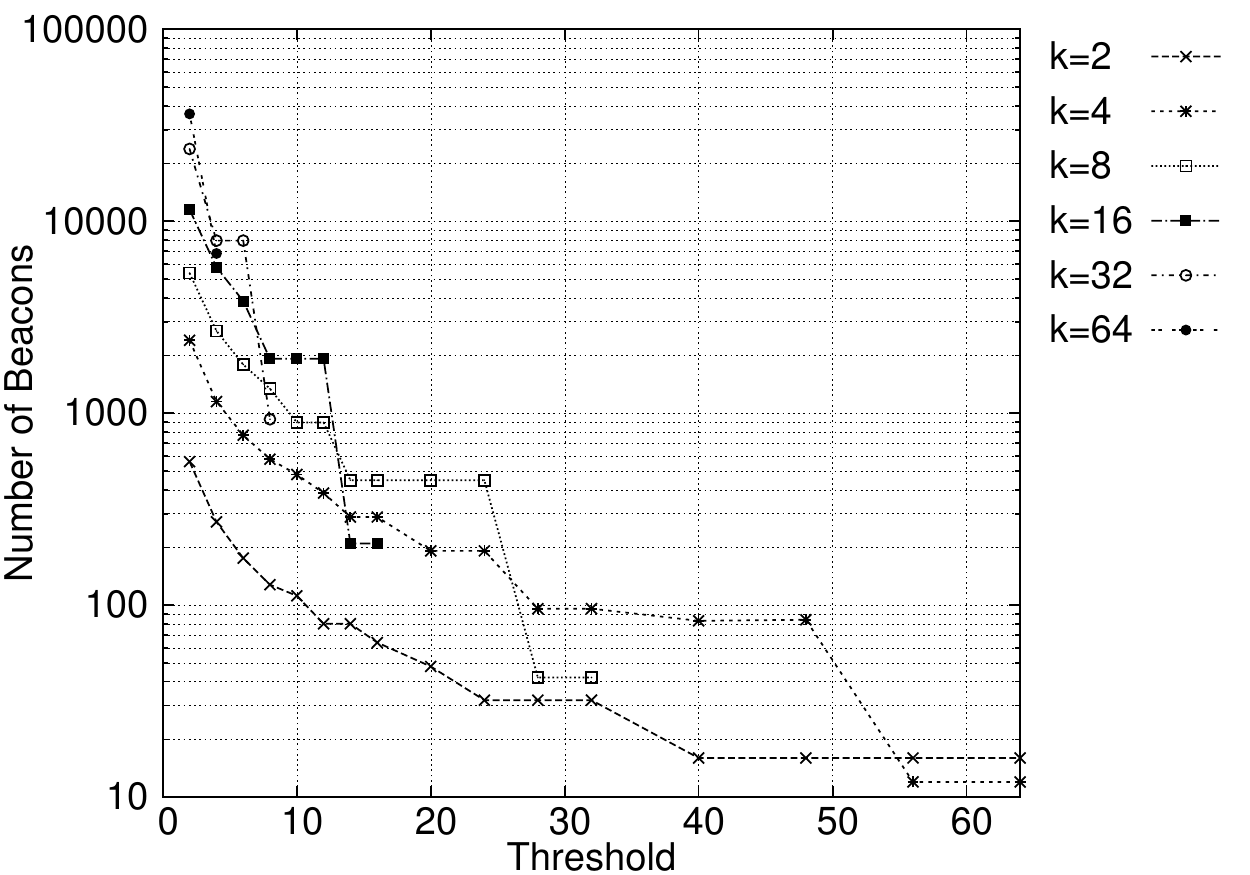}
\captionsetup{width=0.88\textwidth}
\caption{Total number of transmitted beacons for workload status
  communication versus the threshold $\Delta n_{th}$ having different
  numbers of global nodes}
\label{fig:vol}
\end{subfigure}
\caption{Application interference on 256 homogeneous processing elements}
\hrule
\end{figure*}

\subsection{Application Interference}

\begin{figure}
\centering
\includegraphics[width=.36\textwidth]{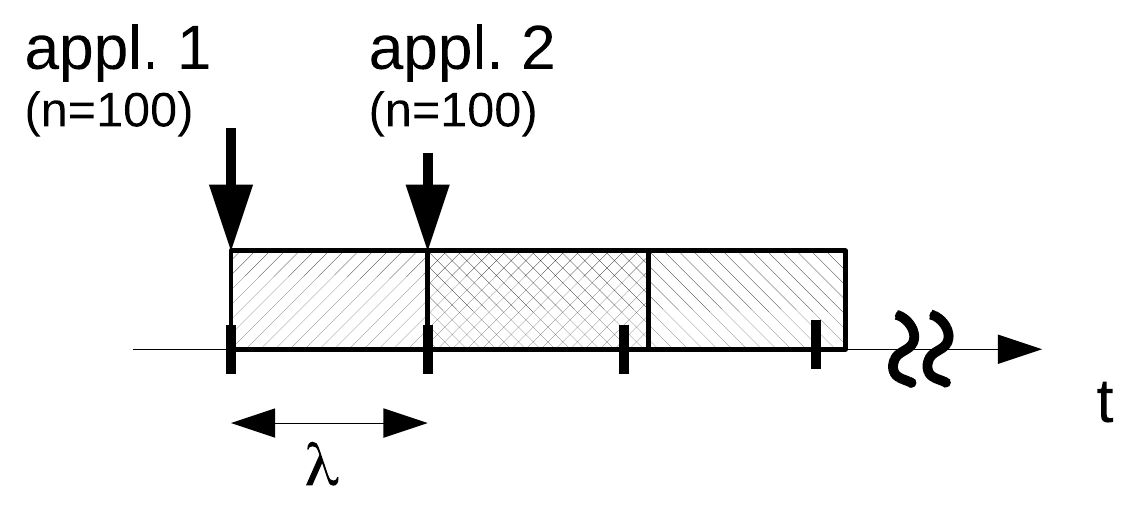}
\caption{Periodic start-up sequence for two competing applications with
  inter-arrival time $\lambda$ and $n$ child tasks}
\label{fig:infr}
\end{figure}

In the second experiment we included interference between two
competing applications having equal priority. The application start-up
sequence with inter-arrival time $\lambda$ as shown in
\mbox{Fig. \ref{fig:infr}} is repeated periodically. The inter-arrival
\mbox{time $\lambda$} is Poisson distributed and has a mean value of
$\lambda=7999$ Ticks. The number of processing elements is $m=256$ and
each application has $n=100$ child tasks. The child task length has a
uniform distribution between \mbox{95 - 100 \%} of the maximum
computation time. The synchronization between the parent and the
child tasks is done by means of the fork/join mechanism presented in
\mbox{Sec. \ref{sec:tman}}.  The stimulus is active for 90 \% of the
simulation time and is send with highest priority directly to a
randomly chosen global node. The other global nodes are kept
\textit{agnostic} about arriving applications and must update their
information according to the presented status communication (see
Sec. \ref{sec:stat}). We do not display any values, where the number
of completed applications differs from the number of injected ones (no
misses are allowed).

Fig. \ref{fig:thr} shows the resulting application speedup for the
hierarchical task mapping algorithm (see Sec. \ref{sec:tman}) and the
threshold-based status communication mechanism. Using $k=16$ global
hardware nodes and a threshold $\Delta n_{th}=4$ a speedup improvement
by a factor of 2.8 compared to $k=1$ is achieved. For a
fully-distributed configuration the improvement factor is only around
1.6 compared to $k=1$. Using the given benchmark, the threshold based
mechanism reveals a robust load balancing as long as the threshold is
smaller than the number of processing elements per cluster. 

Further, we display the number of transmitted status beacons for the
threshold-based mechanism in \mbox{Fig. \ref{fig:vol}}. The figure
gives an indicator about the required energy for status communication,
which is related to the number of \textit{received} beacons. Every
transmitted beacon must be received by all remote nodes to fully
synchronize the network. For a threshold of $\Delta n_{th}=4$ it is
indicated that a fine-grained clustered configuration with $k=32$
management nodes must transmit an amount of beacons that is around
1.37 higher compared to a configuration having \mbox{$k=16$ nodes}.

For a preliminary area analysis we compare an in-house implementation
of a dedicated 32-Bit stack machine as global management node (GMN)
to an mLite/PLASMA CPU \cite{rhoads2006plasma} as processing
element. Both designs have been synthesized using an industrial 65nm
low-power technology (see Tab. \ref{tab:sm}). When disregarding an
additional multiplier having $3547 {\mu m}^2$ shipped inside the
mLite, we still can report around 25\% less area for the stack
machine.

\begin{table}[h]
\centering
\caption{Synthesis results for 65nm low-power}
\label{tab:sm}
\begin{tabular}{|l||r|r|l|}
\hline
Unit   & Comb. [${\mu m}^2$] & Non-comb. [${\mu m}^2$] & $T_{clk}$ \\
\hline\hline
GMN    & 9290.4              & 9881.2                  & 1.77 ns \\
\hline
mLite  & 16268.4             & 12909.5                 & 1.79 ns \\ 
\hline
\end{tabular}
\end{table}

\subsection{Summary} 

In Tab. \ref{tab:sum} we summarize the results of our evaluation in
terms of application speedup using the presented hardware
infrastructure. As a comparison we give our obtained values for a
fully-centralized configuration (like e.g. Nexus++
\cite{dallou2012hardware}) and a fully-distributed one (like
e.g. Isonet \cite{lee2013isonet}). The table indicates the significant
impact of the management overhead, which was constituted by
Eqn. (\ref{eqn:over}), (\ref{eqn:comp}) and (\ref{eqn:sync}). As a
further work, we plan to consider a cycle-accurate model of the task
manager and analyze the overall power consumption of the system. To
get a more realistic scenario about the user applications, their
memory access will be considered as well.

\begin{table}
\centering
\caption{ Comparison of Speedup ($S=t_{seq}/t_{par}$) for $n=100$
  independent tasks on \mbox{$m=256$} PEs using different numbers of
  cluster (global nodes) $k$.  }
\label{tab:sum}
\begin{tabular}{|l||l|p{5cm}|}
\hline
k                              & Speedup & Ref. \tabularnewline
\hline\hline
\parbox[c][1.4em][c]{4mm}{1}   &    28.1 & Centralized, like e.g. Nexus++ \cite{dallou2012hardware} \tabularnewline
\hline
8                              &    73.5 & this work                         \tabularnewline
\hline
16                             &    78.7 & this work                         \tabularnewline
\hline
\parbox[c][1.4em][c]{4mm}{256} &    44.3 & Distributed, like e.g. Isonet \cite{lee2013isonet}       \tabularnewline
\hline
\end{tabular}
\end{table}

\section{Conclusion}
\label{sec:sum}

A dedicated infrastructure of hardware nodes for run-time task
management has been introduced. Compared to previous works we consider
a full-fledged and separated task management infrastructure.  The
infrastructure uses a message passing protocol and allows a design
trade-off between the advantages of centralized and fully-distributed
architectures by choosing an optimal cluster size.

We analyze the clustered architecture by means of an analytic
description as well as by transaction level simulations using a
parallel benchmark including application interference. Our simulations
revealed significant impact of the management overhead to the overall
system performance.

The management overhead for the task mapping problem can be reduced by
using our infrastructure and a two-stage task mapping approach. Having
$m=256$ processing elements and choosing the optimal cluster size can
provide a performance improvement by a factor of 2.8 compared to a
single-cluster/centralized configuration.

The results further show the dependency of the run-time management
system on the status information from remote clusters. The lack of
information may lead to inappropriate mapping decisions causing a
performance drawback. We measured the communication overhead by
counting the number of status beacons transmitted by the global
management nodes. Using a threshold-based mechanism for status
communication and the optimal cluster size, we measured a significant
reduction in terms of transmitted synchronization messages compared to
more fine-grained clustered configurations.

\bibliographystyle{abbrv}
\bibliography{articles,books,misc,proceedings}

\end{document}